\documentclass[aps,preprint]{revtex4}%
\usepackage{amsfonts}
\usepackage{amsmath}
\usepackage{amssymb}
\usepackage{graphicx}%
\providecommand{\U}[1]{\protect\rule{.1in}{.1in}}

\begin{document}
\title{Spin-orbit interaction effects on the magnetoplasmon spectrum of modulated
two-dimensional electron gas}
\author{M. Tahir$^{\ast}$}
\affiliation{Department of Physics, University of Sargodha, Sargodha 40100, Pakistan.}
\author{K. Sabeeh$^{\dag}$}
\affiliation{Department of Physics, Quaid-i-Azam University Islamabad 45320, Pakistan.}


\begin{abstract}
We present a theoretical study of magnetoplasmon spectrum of a two-dimensional
electron gas in the presence of Rashba spin-orbit interaction (RSOI),
one-dimensional weak electric modulation and a perpendicular magnetic field.
The intra-Landau-band magnetoplasmon spectrum is determined in the presence of
spin-orbit interaction within the self consistent field approach at finite
temperature. Due to Rashba effect, the spin of finite-momentum electrons feels
a magnetic field perpendicular to the electron momentum in the inversion
plane. The magnetoplasmon spectrum of the modulated two-dimensional electron
gas (M2DEG) system is found to exhibit beating of Weiss oscillations due to
Rashba effect which is the focus of this work. This effect is absent in the
magnetoplasmon spectrum of M2DEG if Rashba spin-orbit interaction is not taken
into account. In addition, our finite temperature theory ficilitates analysis
of effects of temperature on the magnetoplasmon spectrum of M2DEG in the
presence of RSOI. We find that the beating pattern is damped but continues to
persist at a finite but low temperature.

\end{abstract}
\date[Date text]{date}
\received[Received text]{date}

\revised[Revised text]{date}

\accepted[Accepted text]{date}

\published[Published text]{date}

\maketitle

\section{Introduction}

There continues to be a great deal of interest in the investigation and
manipulation of the spin degree of freedom of the charge carriers in
semiconductor heterostructures with possible applications in the emerging
field of spintronics\cite{1}. Spintronics is based on manipulation of the spin
degree of freedom of the carriers in order to develop functional mesoscopic
devices with wide ranging applications, such as spin filters\cite{2}, spin
field effect transistors\cite{3}, field effect switches\cite{4}, data
storage\cite{5}, quantum computing\cite{6} and even biological sensors\cite{7}%
. The manipulation of the spin degree of freedom requires that we have control
over the strength of spin-orbit interaction that couples the orbital motion of
the electrons with the orientation of electron spins. It was first proposed by
Bychkov and Rashba\cite{8} that in the presence of Rashba spin-orbit
interaction (RSOI), that arises as a result of structural inversion asymmetry,
the spin of finite momentum electrons feels a magnetic field perpendicular to
the electron momentum in the inversion plane. Both experimental as well as
theoretical studies have shown that RSOI has an important role in
semiconductor spintronic systems\cite{1}, since the strength of RSOI can be
controlled by applying a gate voltage on top of the two-dimensional electron
gas (2DEG)\cite{9}.

In addition to the investigation of single particle properties in the presence
of RSOI, there is also serious effort to examine the effects of RSOI on the
collective properties of semiconductor heterostructures. Single particle
magneto-oscillatory phenomena such as the Shubnikov-de Haas and de Haas-van
Alphen effects have been important in probing the electronic structure of
solids. Their collective analog yields important insights into collective
phenomena. In a 2DEG, collective excitations are induced by electron-electron
interactions. These collective excitations (plasmons) are among the most
important electronic properties of a system. In the presence of an external
magnetic field, these collective excitations are known as magnetoplasmons.
Magnetic oscillations of the plasmon frequency occur in a magnetic field. In
this regard, there have been recent studies of collective excitations
(plasmons) of two-dimensional electron gas (2DEG) systems realized in
semiconductor heterostructures in the presence of RSOI with and without an
external magnetic field \cite{10,11,12,13}. These studies show that the
presence of RSOI mixes the spin up and spin down states of neighboring Landau
levels with the result that two new, unequally spaced branches arise.

Given the importance RSOI has acquired, it is important to investigate the
effects of RSOI on the magnetoplasmons of a modulated 2DEG. In the absence of
RSOI but as a result of modulation, Weiss\cite{14,15,16} oscillations in the
magnetoplasmon spectrum appear\cite{17}. These oscillations are due to the
commensurability of two characteristic length scales of the system: the
cyclotron diameter at the Fermi level and the period of the electric
modulation. Weiss oscillations are distinctly different from the Shubnikov-de
Hass (SdH)\cite{18} oscillations that appear at higher magnetic field
strengths. The period of Weiss oscillations depends on both the modulation
period and the square root of the number density of the 2DEG, in contrast to
the linear dependence on the number density of the SdH oscillations. Moreover,
the amplitude of Weiss oscillations is weakly affected by temperature as
compared to SdH oscillations. Theoretical study of RSOI effects on single
particle quantum transport in a modulated 2DEG was recently carried
out\cite{19}, where modulation leads to the broadening of the levels of two
unequally spaced branches into bands whose widths oscillate as a function of
magnetic field. Furthermore, recently RSOI effects on magnetoplasmon
excitations without modulation have been addressed\cite{13} but RSOI effects
on commensurability oscillations in magnetoplasmons of a M2DEG have not been
discussed so far. This is the subject of the present paper. The present work
examines RSOI effects on the collective excitations (magnetoplasmons) of a
modulated 2DEG. This investigation is performed within the
self-consistent-field approach at finite temperature. We analyze the dynamic,
nonlocal dielectric function of the system and highlight the
modulation-induced effects on the intra-Landau band magnetoplasmon spectrum in
the presence of RSOI. In addition, our finite temperature theory facilitates
analysis of effects of temperature on the magnetoplasmon spectrum of a M2DEG
in the presence of RSOI as there is an apparent gap in the literature\cite{17}
regarding finite temperature effects on the magnetoplasmon spectrum of a M2DEG
even in the absence of RSOI. We expect experimental studies such as in-elastic
light scattering\cite{20} and spectroscopic\cite{21} measurements of
magnetoplasmons spectrum of a modulated system in the presence of RSOI would
allow verification of the model presented here and would be quite revealing as
they pertain to spin-orbit interaction effects on the many-body properties of
a 2DEG.

\section{Formulation}

Our system is a modulated two-dimensional electron gas (M2DEG) in the presence
of \ an external magnetic field and RSOI. The magnetic field is applied
perpendicular to the $x-y$ plane in which electrons with unmodulated areal
density $n_{D}$, effective mass $m^{\ast}$ and charge $-e$ are confined. The
Hamiltonian of this system is written as $H=H_{0}+H^{\prime}$, where $H_{0}%
$\ is the unmodulated and $H^{\prime}$ the modulation term. We employ the
Landau gauge and write the vector potential as $\boldsymbol{A}=(0,Bx,0)$. The
two-dimensional Schrodinger equation in the Landau gauge is\cite{13,19};
\begin{equation}
H_{0}=\frac{(\overset{\wedge}{\boldsymbol{p}}+e\overset{\wedge}{\boldsymbol{A}%
})^{2}}{2m^{\ast}}+\frac{\alpha}{\hslash}[\overset{\wedge}{\sigma}%
\times(\overset{\wedge}{\boldsymbol{p}}+e\overset{\wedge}{\boldsymbol{A}}%
]_{z}+\frac{1}{2}g\mu_{B}B\sigma_{z}, \label{1}%
\end{equation}
where $\overset{\wedge}{\boldsymbol{p}}$ is the momentum operator of the
electrons, $g$ the Zeeman factor, $\mu_{B}$\ the Bohr magneton, $\overset
{\wedge}{\sigma}=(\sigma_{x},\sigma_{y},\sigma_{z})$\ the Pauli spin matrices,
$\alpha$\ the strength of the RSOI with%

\begin{equation}
H^{\prime}=V_{0}\cos(Kx). \label{2}%
\end{equation}
$K=2\pi/a$, with $a$ the period of modulation and $V_{0}$ the amplitude of
modulation. Since the Hamiltonian does not depend on the $y$ coordinate, the
unperturbed wavefunctions are plane waves in the $y$-direction. This allows us
to write for the wavefunctions,
\begin{equation}
\phi_{nk_{y}}(\bar{x})=\frac{1}{\sqrt{L_{y}}}e^{ik_{y}y}\sum_{n=0}^{\infty
}u_{n,k_{y}}(x+x_{0})\binom{C_{n}^{+}}{C_{n}^{-}}, \label{3}%
\end{equation}
with $L_{y}$ being the normalization length in the $y$-direction and $\bar{x}$
a 2D position vector on the $x$-$y$ plane, $x_{_{0}}=l^{2}k_{y}=\frac{\hslash
k_{y}}{m^{\ast}\omega_{c}},$ is the coordinate of cyclotron orbit center and
$l=\sqrt{\frac{\hslash}{eB}}$ is the magnetic length, $m^{\ast}$ is the
effective mass. In the $x$-direction, the Hamiltonian has the form of a
harmonic oscillator Hamiltonian. Hence, we can write the unmodulated
eigenstates in the form $\phi_{nk_{y}}(\bar{x})=\frac{1}{\sqrt{L_{y}}%
}e^{ik_{y}y}u_{nk_{y}}(x;x_{0}),$ with $u_{nk_{y}}(x;x_{0})=(\sqrt{\pi}%
2^{n}n!l)^{\frac{-1}{2}}\exp(-\frac{1}{2l^{2}}(x-x_{0})^{2})H_{n}%
(\frac{x-x_{0}}{l}),$ where $u_{nk_{y}}(x;x_{0})$ is a normalized harmonic
oscillator wavefunction centered at $x_{0}$ and $H_{n}(x)$ are Hermite
polynomials with $n$ the Landau level quantum number. The energy of the lowest
Landau level is
\begin{equation}
\varepsilon_{0}^{+}=\varepsilon_{0}=1/2\hslash\omega_{c}-g\mu_{B}B/2 \label{4}%
\end{equation}
with wave function $\phi_{0}^{+}(k_{y})=\frac{1}{\sqrt{L_{y}}}e^{ik_{y}y}%
u_{0}(x+x_{0})\binom{0}{1}.$ For $n=1,2,3,....$, there are two branches of
levels, denoted by $+$ and $-$, with energies%

\begin{equation}
\varepsilon_{n}^{\pm}=n\hslash\omega_{c}\pm\lbrack\varepsilon_{0}^{2}%
+2n\alpha^{2}/l^{2}]^{1/2}. \label{5}%
\end{equation}
The $+$ branch is described by the wave function $\phi_{n}^{+}(k_{y})=\frac
{1}{\sqrt{L_{y}A_{n}}}e^{ik_{y}y}\binom{D_{n}u_{n-1}(x+x_{0})}{u_{n}(x+x_{0}%
)}$\ and $-$ one by \ $\phi_{n}^{-}(k_{y})=\frac{1}{\sqrt{L_{y}A_{n}}%
}e^{ik_{y}y}\binom{u_{n-1}(x+x_{0})}{-D_{n}u_{n}(x+x_{0})}$, where
$A_{n}=1+D_{n}^{2}$ and $D_{n}=(\sqrt{2n}\alpha/l)/[\sqrt{\varepsilon_{0}%
^{2}+2n\alpha^{2}/l^{2}}].$

In the presence of modulation, the Hamiltonian is augmented by the term
$H^{\prime}.$ In this work, we consider the modulation to be weak such that it
is about an order of magnitude smaller than Fermi energy ($V_{0}\ll
\varepsilon_{F}$). Hence, we may employ first order (in $H^{\prime})$
perturbation theory in the evaluation of the energy eigenvalues. The energy
eigenvalues in the presence of modulation and taking into account RSOI
are\cite{19}%

\begin{equation}
\varepsilon^{\pm}(n,x_{0})=\varepsilon_{n}^{\pm}+V_{n}^{\pm}\cos(Kx_{0}),
\label{6}%
\end{equation}
where $V_{n}^{+}=V_{0}e^{-u/2}[D_{n}^{2}L_{n-1}(u)+L_{n}(u)]/A_{n},$
$V_{n}^{-}=V_{0}e^{-u/2}[L_{n-1}(u)+D_{n}^{2}L_{n}(u)]/A_{n},u=\frac
{K^{2}l^{2}}{2}=(\frac{2\pi}{a})^{2}\frac{\hslash}{2m^{\ast}\omega_{c}},$and
$L_{n}(u)$ is a Laguerre polynomial. The above equation shows that the
formerly sharp Landau levels are now broadened into minibands by the
modulation potential. Furthermore, the Landau bandwidths oscillate as a
function of $n$, since $L_{n}(u)$ is an oscillatory function of its index.
These Landau bands become flat for different values of $B$. Flat bands occur
for those values of $B$ for which modulation strength becomes zero. By putting
$V_{n}^{\pm}=\exp(-\frac{u}{2})[L_{n}(u)+L_{n-1}(u)]=0$ one can get the flat
band condition using the asymptotic expression\cite{19}%

\begin{equation}
\exp(-\frac{u}{2})L_{n}(u)\simeq\frac{1}{\sqrt{\pi\sqrt{nu}}}\cos(2\sqrt
{nu}-\frac{\pi}{4}), \label{7}%
\end{equation}
$L_{n}(u)=L_{n-1}(u),$and $2\sqrt{u}\{\sqrt{\pi n_{D}}l\mp\frac{\alpha}%
{\sqrt{2}\hslash\omega_{c}l}\}=\pi(i-1/4),$with $n_{D}$ being the electron
density, one obtains the following condition
\begin{equation}
2R_{c}^{\pm}=a(i-1/4),\text{\ \ \ }i=1,2,3,.......... \label{8}%
\end{equation}
with $R_{c}^{\pm}=R_{c}^{0}\mp\frac{\alpha}{\hslash\omega_{c}}$, $R_{c}^{0}$
the cyclotron radius without RSOI, the upper and lower sign corresponds to the
$\pm$\ branch, $R_{c}^{\pm}=l\sqrt{2n^{\pm}+1},$ is the classical cyclotron
orbit. From Eq. $(7)$ it can be observed that, in the large $n$ limit electron
bandwidth oscillates sinusoidally and is periodic in $1/B,$ for fixed values
of $n$ and $a.$When $n$ is small bandwidth still oscillates, but the condition
$(7)$ no longer holds because neither eq. $(7)$ nor $L_{n}(\chi)\simeq
L_{n-1}(\chi)$ is valid. Interestingly, for low values of $B$, when many
Landau levels are filled, both the systems have the same flat band
condition\cite{19}.

The dynamic and static response properties of an electron system are all
embodied in the structure of the density-density correlation function. We
employ the Ehrenreich-Cohen Self-Consistent Field (SCF) approach\cite{22} to
calculate the density-density correlation function. The SCF treatment
presented here is by its nature a high density approximation which has been
successful in the study of collective excitations in low-dimensional
systems\cite{10,11,12,13,17,23,24} (semiconductor superlattices and quantum
wire structures). Such success has been convincingly attested by the excellent
agreement of SCF predictions of plasmon spectra with experiments.

Following the SCF approach, the density-density correlation function of the
interacting system can be expressed as%
\begin{equation}
\Pi^{\pm}(\bar{q},\omega)=\frac{\Pi_{0}^{\pm}(\bar{q},\omega)}{1-v_{c}(\bar
{q})\Pi_{0}^{\pm}(\bar{q},\omega)} \label{9}%
\end{equation}
with $\Pi_{0}^{\pm}(\bar{q},\omega)$ the density-density correlation function
of the non-interacting system, $v_{c}(\bar{q})=$ $\frac{2\pi e^{2}}{k\bar{q}}$
the 2-D Coulomb potential, $k$ being the dielectric constant and $\bar{q}$ is
the two-dimensional wave number. Making use of the transformation
$k_{y}\rightarrow-k_{y}$, with the fact that $\varepsilon^{\pm}(n,k_{y})$ is
an even function of $k_{y},$ and at the same time interchanging
$n\leftrightarrow n^{\prime}$ we write for the non-interacting density-density
correlation function appearing in equation (9)
\begin{align}
\Pi_{0}^{\pm}(\bar{q},\omega)  &  =\frac{2m^{\ast}\omega_{c}}{\pi\hslash
a}\sum C_{nn^{\prime}}(\frac{l^{2}\bar{q}^{2}}{2})\int\limits_{0}^{a}%
dx_{0}[f^{\pm}(\varepsilon(n^{\prime},x_{0}+x_{0}^{\prime})-f^{\pm
}(\varepsilon(n,x_{0}))]\nonumber\\
&  \times\lbrack\varepsilon^{\pm}(n^{\prime},x_{0}+x_{0}^{\prime}%
)-\varepsilon^{\pm}(n,x_{0})+\hslash\omega+i\eta]^{-1}. \label{10}%
\end{align}
In writing the above equation we converted the $k_{y}$-sum into an integral
over $x_{0}$. $f(\varepsilon(n,x_{0}))$ is the Fermi-Dirac distribution
function, $C_{nn^{\prime}}(x)=\frac{n_{2}!}{n_{1}^{\prime}!}e^{-x}%
x^{n_{1}-n_{2}}[L_{n_{2}}^{n_{1}-n_{2}}(x)]^{2}$with $x=\frac{l^{2}\bar{q}%
^{2}}{2}$ , $n_{1}$= max($n$, $n^{\prime}$), $n_{2}$= min($n$, $n^{\prime}$),
and $L_{n}(x)$ an associated Laguerre polynomial $x_{0}=-\frac{\hslash k_{y}%
}{m^{\ast}\omega_{c}},$ and $x_{0}^{\prime}=-\frac{\hslash q_{y}}{m^{\ast
}\omega_{c}}$. Without modulation ($V_{n}^{\pm}=0$), the above expression
reduces to the result in \cite{13}.

The above equations (9, 10) will be the starting point of our examination of
the intra-Landau band plasmons. These correlation functions are the essential
ingredients for theoretical considerations of such diverse problems as high
frequency and steady state transport, static and dynamic screening and
correlation phenomena.

\section{Intra-Landau-Band plasmon spectrum in the presence of RSOI at finite
temperature}

The plasma modes are readily furnished by the roots of the longitudinal
plasmon dispersion relation obtained from equation (9) as
\begin{equation}
1-v_{c}(\bar{q})\operatorname{Re}\Pi_{0}^{\pm}(\bar{q},\omega)=0 \label{11}%
\end{equation}
along with the condition $Im\Pi_{0}^{\pm}(\bar{q},\omega)=0$ to ensure
long-lived excitations. Equation (11) can be expressed as%
\begin{equation}
1=\frac{2\pi e^{2}}{k\bar{q}\hslash}\frac{2m^{\ast}\omega_{c}}{\pi a}%
\underset{n,n^{\prime}}{\sum}C_{nn^{\prime}}(\frac{l^{2}\bar{q}^{2}}%
{2})(I^{\pm}(\omega)+I^{\pm}(-\omega)), \label{12}%
\end{equation}
with
\begin{equation}
I^{\pm}(\omega)=P\int\limits_{0}^{a}dx_{0}\frac{f^{\pm}(\varepsilon(n,x_{0}%
))}{\hslash\omega-\varepsilon^{\pm}(n,x_{0})+\varepsilon^{\pm}(n^{\prime
},x_{0}+x_{0}^{\prime})}, \label{13}%
\end{equation}
where $P$ is the principal value.

The magnetoplasmon modes originate from two kinds of electronic transitions,
those involving different Landau bands (inter-Landau band
plasmons)\cite{10,11,12,13} and those within a single Landau-band
(intra-Landau band plasmons). Inter-Landau band plasmons involve the local 1D
magnetoplasma mode and the Bernstein-like plasma resonances\cite{25}, all of
which involve excitation frequencies greater than the Landau-band separation
($\sim\hslash\omega_{c}$). On the other hand, intra-Landau band
magnetoplasmons resonate at frequencies comparable to the bandwidths, and the
existence of this new class of modes is due to finite width of the Landau
levels caused by the modulation. In order to investigate the effects of RSOI
and the modulation on the magnetoplasmon spectrum requires the study of the
intra-Landau band magnetoplasmons. In the absence of modulation, the plasmon
spectrum of a 2DEG with RSOI has been analyzed in detail
elsewhere\cite{10,11,12,13} with and without an external magnetic field. In
the present work, we investigate the magnetoplason spectrum with RSOI in the
presence of modulation. We show below that as a result of both modulation and
RSOI, Weiss oscillations in the magnetoplasmon spectrum are found to exhibit a
beating pattern. This occurs in the low magnetic field regime. At higher
fields, Weiss oscillations are suppressed and Shubnikov de Haas (SdH)
oscillations are the dominant magnetic oscillations. SdH type of oscillations
result from the emptying out of electrons from successive Landau bands when
they pass through the Fermi level as the magnetic field is increased for a
fixed value of RSOI. The amplitude of the SdH type of oscillations is a
monotonic function of magnetic field, when the Landau bandwidth is independent
of the band index $n$. In a M2DEG in the presence of RSOI considered here, the
Landau bandwidths oscillate as a function of the band index $n.$which
significantly affects the plasmon spectrum of the intra-Landau band type.

For the excitation spectrum, we need to numerically solve equation (12) for
all vectors, energies, periodic modulation strength, temperature and magnetic
field. We will consider the case of weak modulation ($V_{0}/E_{F}<<1$),
constant RSOI strength and long wave length. In this case we are concerned
with transitions within a Landau miniband, i.e. $n=n^{\prime}$, $\varepsilon
_{n^{\prime}}^{\pm}-\varepsilon_{n}^{\pm}=0$ and $C_{nn^{\prime}%
}(x)\rightarrow1$\cite{13}. Hence, we are able to solve equation (12)
analytically at finite temperature.

The intra-Landau-band plasmon dispersion relation at finite temperature
reduces to $1=\frac{\overset{\sim}{\omega}^{2}}{\omega^{2}}$, where
\begin{equation}
\hslash^{2}\overset{\sim}{\omega}^{2}=\frac{16e^{2}}{\pi k\bar{q}\hslash}%
\frac{m^{\ast}\omega_{c}}{a}\times\{\sin^{2}(\frac{\pi}{a}(x_{0}^{\prime
})F_{n}^{\pm}(u)\}, \label{14}%
\end{equation}%
\begin{equation}
F^{\pm}(u)=\underset{n}{-\sum}V_{n}^{\pm}\times\int_{0}^{a/2}dx_{0}f^{\pm
}(\varepsilon(n,x_{0}))\cos(Kx_{0}) \label{15}%
\end{equation}
and $f^{\pm}(\varepsilon(n,x_{0}))$\ is the distribution function. Without
RSOI ($\alpha=0$) and at zero temperature, this expression reduces to the
result obtained in \cite{17}. In the regime of weak modulation%

\begin{equation}
f^{\pm}(\varepsilon(n,x_{0}))\simeq f^{\pm}(\varepsilon_{n})+f^{\pm\prime
}(\varepsilon_{n})\{V_{n}^{\pm}\cos(Kx_{0})\}, \label{16}%
\end{equation}
where $f^{\prime}(x)=\frac{d}{dx}f(x)$ is the derivative of the Fermi Dirac
distribution function. After the substitution of this expansion in equations
(14 \& 15) and performing the integral over $x_{0}$, the intra-Landau band
spectrum is obtained%
\begin{equation}
\hslash^{2}\overset{\sim}{\omega}^{2}=\frac{4e^{2}m^{\ast}\omega_{c}}{\pi
k\bar{q}\hslash}\times\{\sin^{2}[\frac{\pi}{a}(x_{0}^{\prime})]\times\lbrack
G^{++}+G^{--}]\}, \label{17}%
\end{equation}
here%

\[
G^{++}=\underset{n}{\sum}\left\vert V_{n}^{+}\right\vert ^{2}\times
\lbrack-f^{+\prime}(\varepsilon_{n})]=\underset{n}{\sum}\left\vert V_{n}%
^{+}\right\vert ^{2}\times\beta\lbrack f^{+}(\varepsilon_{n})\{1-f^{+}%
(\varepsilon_{n})\}],
\]
and
\[
G^{--}=\underset{n}{\sum}\left\vert V_{n}^{-}\right\vert ^{2}\times
\lbrack-f^{-\prime}(\varepsilon_{n})]=\underset{n}{\sum}\left\vert V_{n}%
^{-}\right\vert ^{2}\times\beta\lbrack f^{-}(\varepsilon_{n})\{1-f^{-}%
(\varepsilon_{n})\}]\ ,
\]
where $\beta$=$K_{B}T$ with $K_{B}$\ being the Boltzmann constant. Since we
have $\Pi^{--}$ and $\Pi^{++},$ this term leads to an extra negative term for
the total intraband correlation\cite{10,11} $\Pi^{++}+\Pi^{--}$. The interband
correlation is negligible in the long wave length limit because electron spins
at the same wave vector in different branches are opposite.

We have derived the expression for $\hslash\overset{\sim}{\omega}$ (equation
17) under the condition $\hslash\omega>>\mid\varepsilon^{\pm}(n,x_{0}%
+x_{0}^{\prime})-\varepsilon^{\pm}(n,x_{0})\mid$ as $x_{0}^{\prime}%
\rightarrow0$ which leads to a relation between the energy and the Landau
level broadening $\hslash\omega>>\mid2\left\vert V_{n}^{\pm}\right\vert
\{\sin(\frac{\pi}{a_{x}}x_{0}^{\prime})\sin[(\frac{2\pi}{a_{x}})(x_{0}%
+\frac{x_{0}^{\prime}}{2})]\}\mid$. This ensures that $Im\Pi_{0}(\bar
{q},\omega)=0$ and the intra-Landau-band magnetoplasmons are undamped. For a
given $\left\vert V_{n}^{\pm}\right\vert $, this can be achieved with a small
but nonzero $q_{y}$ (recall that $x_{0}^{\prime}=-\frac{\hslash q_{y}}%
{m^{\ast}\omega_{c}}$). In general, the inter- and intra-Landau-band modes are
coupled for arbitrary magnetic field strengths. Only the intra-Landau-band
mode ( $\hslash\overset{\sim}{\omega})$ will be excited in the energy regime
$\hslash\omega_{c}>\hslash\omega\sim$ $\left\vert V_{n}^{\pm}\right\vert .$

\section{Discussion of Results}

The intra-Landau-band plasmon energy obtained in equation (17) is shown
graphically in Figs (1,2,3) as a function of $1/B$ for different values of the
temperature (T) and RSOI parameter : $\alpha=0,$ and $\alpha=1.2\times
\alpha_{0}$. The parameters used in all of our figures are
\cite{13,14,15,16,17,19}: $\alpha_{0}=1\times10^{-11}$ eVm, $m^{\ast
}=0.05m_{e}$, $g=2$, $\mu_{B}=5.78\times10^{-5}$ eV/Tesla$,$ $k=14.5$,
$n_{D}=3.16\times10^{-15}$ m$^{-2}$, $a=380$ nm,\ and $V_{0}$ = 0.5 meV. We
also take $q_{x}=0$ and $q_{y}=0.01k_{F},$ with $k_{F}=(2\pi n_{D})^{1/2}$
being the Fermi wave number of the unmodulated 2DEG in the absence of magnetic
field. Furthermore, in Fig (3) the intra-Landau-band plasma energy is shown as
a function of $1/B$ for RSOI strength ($\alpha=1.2\times\alpha_{0}$),
temperature T=1K and all other parameters are the same as in Fig. (2). The
beating of Weiss oscillations is solely due to + and - branch of the banwidth
of the intra-landau band plasmon energy satisfying the condition given in Eq.
(8). In Fig. (1), we show the intra-Landau band magnetoplasmon spectrum in the
absence of spin-orbit interaction $\alpha=0$ at two different temperatures
T=0.25 K, $3.0$ K. The minima of Weiss oscillations in Fig. (1) satisfy the
flat band condition obtained in Eq. (8) for zero Rashba coupling strength. In
this figure, the SdH oscillations are washed out completely at 3K but Weiss
persist and are undamped, which confirm the weak dependence of Weiss
oscillations on temperature compared to SdH oscillations. To highlight the
effects of RSOI, Figs.(2,3) should be compared with Fig.(1). As a result of
finite RSOI, at low magnetic fields (correspondingly higher $1/B$ values) the
Weiss oscillations in the intra-Landau band magnetoplasmon spectrum show
beating pattern. The minima in Fig. (2 \& 3) indicate the individual beats
which follow the flat band condition obtained in Eq.(8) with finite value of
RSOI strength. This occurs due to the mixing of the spin up and spin down
states of neighbouring Landau levels by RSOI resulting in two, unequally
spaced levels. The effect of weak electric modulation potential is to broaden
these levels into bands. The bandwidth of these bands oscillates as a function
of magnetic field. Due to the splitting of the bands, there are two flat band
conditions as opposed to a single condition in the absence of RSOI. As a
result, there are oscillations at two different frequencies leading to the
observed beating pattern. The flat band condition that we arrive at is the
same as obtained in \cite{19} and its implications on the transport
coefficients are discussed in detail there. The origin of beating pattern in
Weiss oscillations can also be understood by a closer analytic examination of
Eq. (17) at zero temperature. In the zero temperature case Eq. (17) can be
expressed as $\hslash\overset{\sim}{\omega}$ $\sim\sqrt{\left[  \left\vert
V_{n}^{+}\right\vert ^{2}\right]  _{\varepsilon_{F}}+\left[  \left\vert
V_{n}^{-}\right\vert ^{2}\right]  _{\varepsilon_{F}}}$, which is a linear
combinations of the two split bandwidths evaluated at the Fermi energy, these
are oscillatory functions. They oscillate at different frequencies and their
linear combination results in the observed beating pattern of Weiss oscillations.

Furthermore, since the flat band condition depends on the strength of RSOI,
the amplitude and phase of the magnetic oscillations (Weiss and SdH) vary with
the strength of RSOI. Fig (3) shows the effect of temperature on the
magnetoplasmon spectrum while taking into account RSOI. As the temperature is
increased from 0.25 K to 1 K, comparing Figs.(2) $\And$ (3), we see that the
beating pattern in Weiss oscillations of the plasmon spectrum is still present
but damped. These results depend on the strength of RSOI as well as the
temperature. Consider Eq. (17) at finite temperature T. The summations over
$n$ can be performed to yield $\hslash\overset{\sim}{\omega}$ $\sim
\underset{n}{\sum}\sqrt{\{\left\vert V_{n}^{+}\right\vert ^{2}\times
\lbrack-f^{+\prime}(\varepsilon_{n})]+\left\vert V_{n}^{-}\right\vert
^{2}\times\lbrack-f^{-\prime}(\varepsilon_{n})]\}}$. Damping of the plasmon
energy with temperature can be explicitly seen if the derivative of the Fermi
functions appearing in the above expression are expressed as follows:
$-f^{\pm\prime}(\varepsilon_{n})=\beta f^{\pm}(\varepsilon_{n})\{1-f^{\pm
}(\varepsilon_{n})\}$ with $\beta=K_{B}T.$

\section{Conclusions}

We have determined the intra-Landau band magnetoplasmon spectrum for a
modulated two dimensional electron gas in the presence of an external magnetic
field and RSOI employing the SCF approach at finite temperature. The
magnetoplasmon spectrum of the modulated two-dimensional electron gas system
is found to exhibit beating of Weiss oscillations due to Rashba effect as a
function of the inverse magnetic field \ In the regime of low magnetic field,
the modulation induced Weiss oscillations show beating pattern whose period is
determined by the period of electric modulation and the strength of RSOI.
Furthermore, we find that the beating pattern is damped but continues to
persist at a finite but low temperature. Experimental study of these results
should be quite revealing as they directly bear on many body properties of
modulated 2DEG in the presence of RSOI.

One of us (K. Sabeeh) would like to acknowledge the support of the Pakistan
Science Foundation (PSF) through project No. C-QU/Phys (129). M. Tahir would
like to acknowledge the support of the Pakistan Higher Education Commission (HEC).

*Present address: Department of Physics, The Blackett Laboratory, Imperial
College London, South Kensington Campus, London SW7 2AZ, United Kingdom, m.tahir06@imperial.ac.uk

$\dag$Electronic address: ksabeeh@qau.edu.pk; kashifsabeeh@hotmail.com

\end{document}